\documentclass[journal,10pt]{IEEEtran}

\usepackage{amsmath, amsthm, amsfonts}
\usepackage{xcolor}
\usepackage{mathtools}
\usepackage{caption}
\usepackage{subcaption}
\usepackage{graphicx}
\usepackage{longtable}
\usepackage{tabularx}
\usepackage{comment}
\usepackage{cite}
\usepackage{color,soul}
\usepackage[hyphens]{url}
\usepackage{multirow}
\usepackage{graphicx}
\usepackage{enumerate}
\usepackage{enumitem}
\usepackage{array}
\usepackage{booktabs} 
\usepackage{subcaption}
\usepackage{algorithm}
\usepackage[noend]{algpseudocode}
\usepackage{bm}
\usepackage{amssymb}
\usepackage{hyperref}
\usepackage{balance}

\newcommand{\abs}[1]{\left\lvert#1\right\rvert}

\theoremstyle{definition}

\theoremstyle{remark}

\title{A Bi-Level Stochastic Game Model for PMU Placement in Power Grid with Cybersecurity Risks}
\author{\IEEEauthorblockN{\textbf{Saptarshi Ghosh}\IEEEauthorrefmark{1}, \textbf{Murali Sankar Venkatraman}\IEEEauthorrefmark{2}, \textbf{Shehab Ahmed}\IEEEauthorrefmark{1},
\textbf{Charalambos Konstantinou}\IEEEauthorrefmark{1}} \\
\IEEEauthorblockA{
\IEEEauthorrefmark{1}CEMSE Division, King Abdullah University of Science and Technology (KAUST)\\
\IEEEauthorrefmark{2}Data Analytics and AI, ENOWA (NEOM) \\
E-mail: \{saptarshi.ghosh, shehab.ahmed, charalambos.konstantinou\}@kaust.edu.sa,\\ murali.venkatraman@neom.com} \\ 
}
\date{}
\begin{document}
\IEEEaftertitletext{\vspace{-2\baselineskip}}

\maketitle

\begin{abstract}
Phasor measurement units (PMUs) provide accurate and high-fidelity measurements in order to monitor the state of the power grid and support various control and planning tasks.  However, PMUs have a high installation cost prohibiting their massive deployment.  Minimizing the number of installed PMUs needs to be achieved  while also maintaining full observability of the network. At the same time, data integrity attacks on PMU measurements can cause mislead power system control and operation routines. In this paper, a bi-level stochastic non-cooperative game-based placement model is proposed for PMU allocation in the presence of cyber-attack risks. In the first level, the protection of individual PMU placed in a network is addressed, while considering the interaction between the grid operator and the attacker with respective resource constraints. In the second level, the attacker observes the placement of the PMUs and compromises them, with the aim of maximizing the state estimation error and reducing the observability of the network. The proposed technique is deployed in the IEEE-9 bus test system. The results demonstrate a 9\% reduction in the cost incurred by the power grid operator for deploying PMUs while considering cyber-risks.
\end{abstract}
\begin{IEEEkeywords}
Cyber-attacks, game-theory, observability, phasor measurement units, state estimation, risk propagation.
\end{IEEEkeywords}

\vspace{-0.1in}
\section{Introduction}
In the past decades, phasor measurement units (PMUs) provided capabilities for intelligent monitoring and control of the electric power grid. PMUs can give precise and real-time data for monitoring the operational status of the system and enhancing its reliability. However, the overall cost of a PMU (including acquisition,
installation, and commissioning) can be quite high. Thus,  cost-benefit studies for the PMU network is of paramount importance. Consequently, research focuses on the ideal strategies for PMUs placement, in an effort to reduce the number of PMUs on a power system while maintaining system observability \cite{8973553}. Besides the benefits of PMUs in terms of system operation and control, due to the information and communication technology (ICT) nature of PMUs, they also introduce cyber-risks as a prominent target for attackers.

Existing literature demonstrates a variety of cyber-attacks against PMUs. For example, PMUs can be rendered inoperable by preventing their measurements from reaching their intended destination using distributed denial-of-service attacks \cite{7523444}. Malicious programs can also be injected into PMUs to arbitrary change their functionality \cite{8586320}. Attacks including false data injection, data, and GPS  spoofing can alter the actual data and time stamps of PMU measurements \cite{9031416,9387555, konstantinou2017gps}. The effect of cyber-risks and random line outages on system observability is addressed in \cite{DING2021107586}. In literature, the cyber-risks are designed as worst case scenarios, but, in practice, attackers are rational entities who can learn optimal strategies.

The problem of optimal PMU placement towards enhancing power system observability can be expressed as a mixed-integer linear program which can be efficiently solved. In contrast, the formulation of PMU placement towards  optimizing the accuracy of state estimation results in a mixed-integer convex program is computationally more challenging. In order to reduce the complexity of the optimization, the authors in  \cite{8810467} propose to loosen the restrictions and solve a convex problem, with a rounding heuristic used to discover a suitable solution. This technique is also utilized in \cite{6632977, 8115294}, and it enables the management of large-scale power systems in \cite{8810467}, although without any guarantee of optimality. In \cite{7950984}, a technique for optimization based on particle swarm optimization is proposed. The population-based search processes, such as swarm optimization, evolutionary algorithms, etc., do not impose any unique restrictions on the optimization problem's features. As a result, they are utilized for putting PMUs with non-classical optimality criteria. Therefore, it can be stated that distributed techniques can simplify the optimal PMU placement problem in order to decrease state estimation errors. 

In this paper, we propose an alternative game-based solution approach that can be implemented distributively, and models the decision makers as rational and selfish. This makes the proposed model ideal to design practical attack scenarios instead of considering the worst case. The contributions of this work are as follows: 
\begin{itemize}
    \item A bi-level stochastic game based placement model is proposed for PMU allocations in
    the presence of an attacker that jointly addresses the issues of reducing cost, maintaining system observability, and reducing state estimation errors.
    \item Unlike existing literature, where attack is designed as the worst case scenario, in this work, the attacker is a rational and selfish entity that designs the stealthy attack over time in response to the action of the grid operator.
    \item The first level of the game-theoretic formulation consists of repeated interaction between the grid operator and attacker. The closed-form solution of the equilibrium strategies is obtained using the recursive representation of the decision making process.
    \item The proposed PMU placement approach is applied to IEEE 9-bus system. It shows an improvement of 9\% in terms of the number of PMUs required in comparison to conventional placement techniques with cyber-risks.
\end{itemize}
The system model is described in Section II, followed by the problem formulation in Section III. The solution approach is discussed in Section IV, while Section V presents the case study. Finally conclusions are drawn in Section VI.

\vspace{-0.5mm}
\section{System Model }
     \begin{figure*}[!t]
         \centering
         \includegraphics[width=0.65\textwidth]{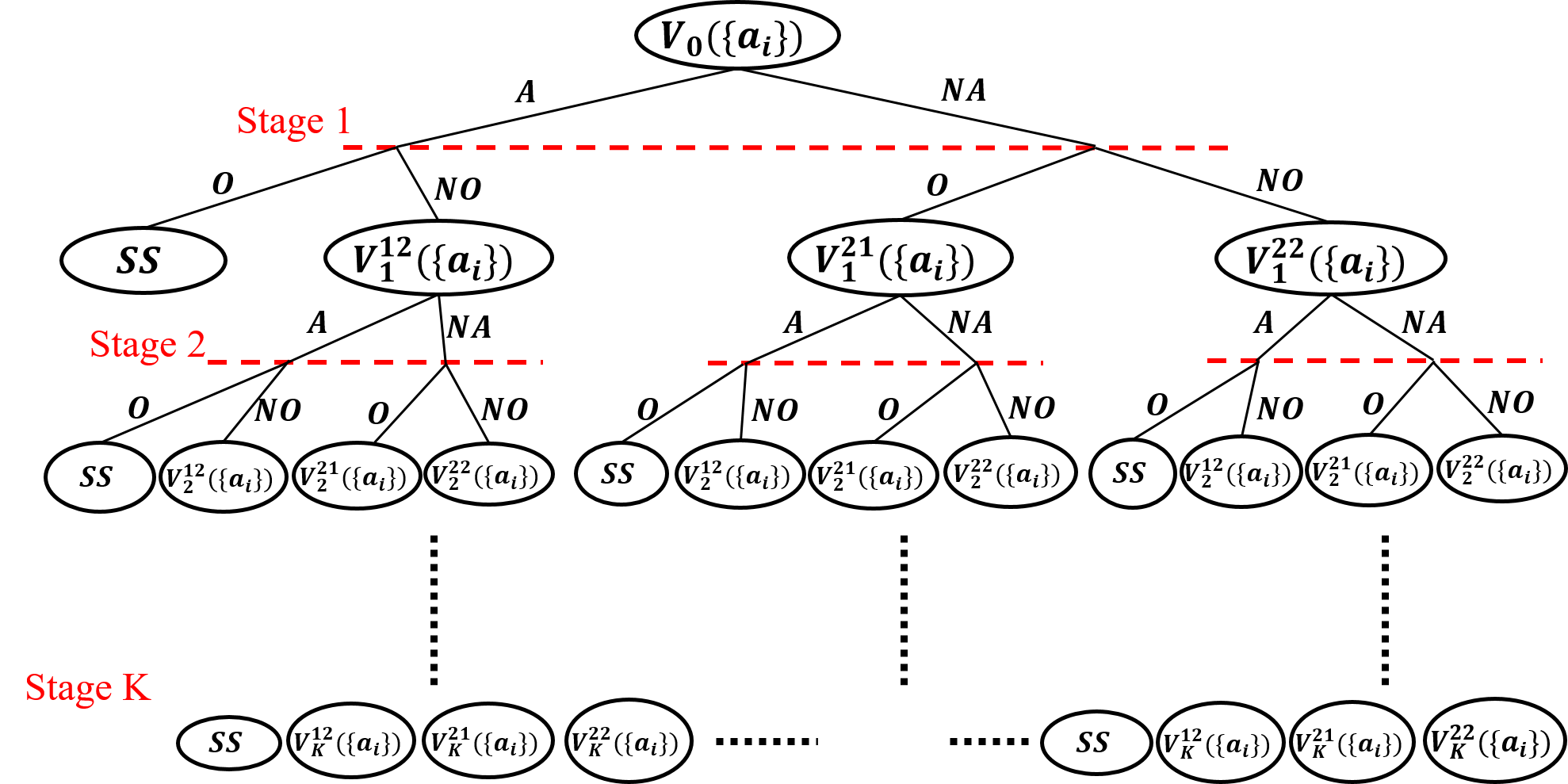}
\caption{Extensive form of representation of grid operator-attacker interaction.}
\vspace{-0.1in}
    \label{game_model}\vspace{-0.1in}
     \end{figure*}
A network of PMUs is considered to monitor the electric power  grid. The PMU measurement vector, denoted as $\mathbf{z}\in \mathbb{R}^{2d}$, includes the real part and imaginary part of the measured bus voltage and branch current phasors, where $d$ is the number of measurements. The entries of the state vector, denoted as $\mathbf{x}\in \mathbb{R}^{2n_b}$, are the real part and imaginary part
of all the bus voltage phasors. $n_b$ is the number of buses. The relationship
between the PMU measurements and the system state variables can be represented as a linear function. The measurement model, considered in this work, is given by:
\begin{equation}\label{2}\mathbf{z}=\boldsymbol{\Gamma}_m \boldsymbol{H}_lx+\mathbf{a}+\boldsymbol{\eta}\end{equation}
where \begin{eqnarray}
\mathbf{H}_l=\begin{bmatrix}\mathbf{I}&\mathbf{0}\\\text{Re}(\mathbf{Y}_l)&-\text{Im}(\mathbf{Y}_l)\\\mathbf{0}&\mathbf{I}\\\text{Im}(\mathbf{Y}_l)&\text{Re}(\mathbf{Y}_l)\end{bmatrix},
\end{eqnarray}
\begin{eqnarray}
\boldsymbol{\Gamma}_m=\begin{bmatrix}
    \gamma^{1} & & \\
    & \ddots & \\
    & & \gamma^{2d}
  \end{bmatrix}.
\end{eqnarray}
The measurement matrix, denoted as 
$\mathbf{H}_l\in \mathbb{R}^{2d\times 2n_b}$, depends on the selected PMU locations, hence the
subscript $l$ denotes a particular combination of PMU deployment. Also, $\mathbf{Y}_l\in \mathbb{R}^{b\times n}$ is the branch admittance matrix, where $b$ is the number of current measurements. $\mathbf{I}\in \mathbb{R}^{g\times g}$ is the identity matrix, where $g$ standing for the
number of PMUs, and the entries of $\mathbf{0}\in\mathbb{R}^{g\times (2n_b-g)}$ are all zeros. $\boldsymbol{\Gamma}_m\in \mathbb{R}^{2d\times 2d}$, $m=1,\dots,2^g$ denotes all $2^g$ possible measurement arrival patterns at each time instant based on the PMU deployment strategy.
The $\mathbf{\Gamma}_m$ is a diagonal matrix having elements $\gamma^i$ denoting the availability of the corresponding measurement, i.e., $\gamma^i=1$, or $0$ otherwise. 
The availability of a measurement $\gamma^i$ depends on whether the PMU generating that measurement  is deployed or not. The decision of the operator to deploy a PMU is given as:
\begin{eqnarray}\setcounter{equation}{1}
y_i=\begin{cases}
0, &\text{PMU not placed at bus}~ i ~\text{with} ~Pr(y_i)\\1, &\text{PMU placed at bus}~ i~ \text{with}~ 1-Pr(y_i)\nonumber
\end{cases}
\end{eqnarray}In case $y_i=1$, i.e., the grid operator placed a PMU at bus \#$i$, the corresponding measurements will be available. $\boldsymbol{\eta} \sim \mathcal{N}(0,\mathbf{R})$ is a $2d\times 1$ additive Gaussian measurement noise vector that is assumed independent across PMUs. 

The maximum likelihood estimate of the system state is:
\begin{eqnarray}\label{state_est}
\hat{\mathbf{x}}=(\sum_{i=1}^{n_b} y_i (\mathbf{H}_i^T \mathbf{R}_i^{-1} \mathbf{H}_i))^{-1}\sum_{i=1}^{n_b} y_i (\mathbf{H}_i^T \mathbf{R}_i^{-1} \mathbf{z}_i)
\end{eqnarray}
The state error covariance is denoted as $y_i(\mathbf{H}_i^T\mathbf{R}_i^{-1}\mathbf{H}_i)$, where $\mathbf{H}_i$ is the submatrix composed of rows of $\mathbf{H}_l$ associated with the PMU measurements at bus \#$i$,  and $\mathbf{R}_i$ is the submatrix of variances of measurement errors $\mathbf{\eta}_i$.
Therefore, for a given strategy of PMU placement $\lbrace y_i\rbrace$, the state error covariance matrix at $\hat{\mathbf{x}}$ is denoted as:
\begin{eqnarray}\label{qa}
\mathbf{G}(\lbrace y_i\rbrace)=\sum_{i=1}^{n_b} y_i (\mathbf{H}_i^T \mathbf{R}_i^{-1} \mathbf{H}_i)
\end{eqnarray} Instead of maximizing/minimizing the covariance matrix in (\ref{qa}), the scalars of the error covariance matrix are considered in this work. The scalars are built by minimizing the variance of the state error (\ref{state_est}). In this regard, the determinant of $\mathbf{G}^{-1}(\lbrace y_i \rbrace)$,
\begin{eqnarray}\label{me}
\boldsymbol{\phi}_D(\lbrace y_i\rbrace)=\text{det}\mathbf{G}^{-1}(\lbrace y_i\rbrace)
\end{eqnarray} is considered as the objective function in this work \cite{8115294}.

The attacks on the PMU network are categorized, in this paper, as direct and indirect. The attacker randomly chooses a PMU or a group of PMUs (denoted as $\lbrace \alpha_i\rbrace$) to launch direct attacks. The probability that PMU $i$ is attacked directly is denoted as:
\begin{eqnarray}
\alpha_i=\begin{cases}
0, &\text{No Attack of PMU at bus}~ i~ \text{with}~ Pr(\alpha_i) \\1, &\text{Attack PMU placed at bus}~ i~ \text{with}~ 1-Pr(\alpha_i)
\end{cases}\nonumber
\end{eqnarray}A PMU, compromised through direct attack, can be used to further propagate the attack to the ``healthy'' PMUs in its neighborhood, resulting in the propagation of risk within the network. This behavior is defined as an indirect attack.  The propagation of this risk in the PMU network can take place over the data communication links among the PMUs \cite{DING2021107586}. Since, the successful data transmission over a link is typically modeled as a random event, the success of an indirect attack on the ``healthy'' PMU is also probabilistic. The probability that a ``healthy'' PMU $j$ is compromised because of the compromised PMU $i$ be denoted by $\beta_{ij}$. The propagation of the risk at time instant $k$ is denoted by $\boldsymbol{\beta}_k=\lbrace \beta_{ij}\rbrace$.  Based on the PMUs deployed by the grid operator and the set of those PMUs compromised by direct or indirect attack, the $2d\times 1$ vector of attack inputs injected by the attacker is denoted by $a$.
The condition of the attack $a$ getting detected is given by:
\begin{eqnarray}\label{det_con}
(\sum_{i=1}^{n_b} y_i \alpha_i(\mathbf{H}_i^T \mathbf{R}_i^{-1} \mathbf{H}_i))^{-1}\sum_{i=1}^{n_b} y_i \alpha_i (\mathbf{H}_i^T \mathbf{R}_i^{-1} a_i)\geq\tau,
\end{eqnarray}
where $\tau$ is the threshold level for bad data detection \cite{cosovic2018distributed}. $\tau$ decides the trade-off in detecting false data and false alarms. The aim of the next section is to optimally decide the PMU placement $\lbrace y_i\rbrace$ in the presence of an attacker who designs the optimal attack $\lbrace \alpha_i\rbrace$ such that the state error covariance is minimized while satisfying the observability constraints.

\section{Problem Formulation}
The problem of deploying the PMUs in presence of direct and direct attacks is modeled as a multi-stage non-cooperative game between the grid operator and the attacker, as shown in Fig. \ref{game_model}. Given the directly attacked set of PMUs $\lbrace \alpha_i\rbrace$, the game can be defined
by a sequence of matrices $\mathbf{S}_1, 
\mathbf{S}_2,\dots,\mathbf{S}_k,\dots, \mathbf{S}_{K}$, where $K$ is the total number of stages. At any stage $k$, when the
grid defender (column player) and attacker (row player) simultaneously
select action $i$ and $j$ from the set $\lbrace \text{Optimize}(O), \text{No Optimize}(NO)\rbrace$
and $\lbrace \text{Attack} (A), \text{No Attack}(NA)\rbrace$, respectively, the attacker payoff
equals $S_k^{ij}$. The payoff $S_k^{ij}$ is calculated by solving the following optimization problem: 

\begin{IEEEeqnarray}{lCr}\label{opt}
&S_k^{O,A}=\min_{\lbrace x_k \rbrace, \lbrace u_{ik} \rbrace} \boldsymbol{\phi}_D(\lbrace y_i\rbrace) \end{IEEEeqnarray}
\begin{IEEEeqnarray}{lCr}\label{opt1}
&\text{s.t.}~ f_i\geq1, z_i=1,~\forall i
\end{IEEEeqnarray}
\begin{IEEEeqnarray}{lCr}\label{opt2}
&Pr(\lbrace \alpha_i\rbrace)=\arg\max_{\lbrace \alpha_k\rbrace}\nonumber\end{IEEEeqnarray}
\begin{IEEEeqnarray}{lCr}\label{opt3}& \sum_{i=1}^{n_b}(\sum_{i=1}^{n_b} \big(y_i \alpha_i(\mathbf{H}_i^T \mathbf{R}_i^{-1} \mathbf{H}_i))^{-1} y_i \alpha_i \mathbf{H}_i^T \mathbf{R}_i^{-1} a_i\big)^2\end{IEEEeqnarray}
\begin{IEEEeqnarray}{lCr}\label{opt4}
&\text{s.t.}~ (\ref{det_con}).
\end{IEEEeqnarray}The buses with loads will be referred to as ``load''
buses, whereas the buses, without any loads, will be
called ``zero-injection'' buses. Based on the deployed network of PMUs, observability constraints for ``load" bus \#$i$ is given as:
\begin{eqnarray}
f_i=\sum_{k=1}^{n_b} a_{ik}\alpha_k x_k+\sum_{k=1}^{n_b}a_{ik}\alpha_k s_iu_{ik}\geq1.
\end{eqnarray} $a_{ij}=1$ denotes whether the bus \#$i$ and \#$j$ are connected. $s_i=1$ denotes that bus \#$i$ is a zero-injection bus. $x_k=1$ denotes that bus \#$k$ has a PMU. $u_{ik}=1$ denotes that PMU on bus \#$k$ can monitor zero-injection bus \#$i$. In case \#$i$ is a zero-injection bus, the observability constraint is: 
\begin{eqnarray}
z_i=\sum_{k=1}^{n_b}a_{ik}\alpha_k u_{ik}=1.
\end{eqnarray}

Based on the probabilities of the direct and indirect attack, the probabilistic observability constraint for bus \#$i$ is: 
\begin{eqnarray}
Pr(f_i\geq 1)=\sum_{i=1}^{n_b}\big(1-Pr(\alpha_i)\prod_{j=1,j\neq i}^{n_b} \beta_{ij}\big).\nonumber
\end{eqnarray} The aim of the attacker is to maximize the distortion in the state estimation process, depicted in (\ref{opt3}), while remaining undetected, shown in the constraint of (\ref{opt4}). Therefore, the attacker either gets detected or remains undetected at stage
$k$ whenever the action pair $\lbrace\text{Attack}, \text{Optimize}\rbrace$ gets selected. The game stops at stage $k$, if the attacker gets detected, or successfully injects error, known as the stopping state (SS). The condition of the SS at the $k^{th}$ stage is formally given as:
\begin{eqnarray}
I(i_k,j_k)=\begin{dcases}1, & \lbrace i_k,j_k \rbrace=\lbrace A, O\rbrace,\\0, & \text{otherwise}.\end{dcases}
\end{eqnarray}

Given a sequence of grid operator and attacker actions $\lbrace (i_1,j_1), \dots, (i_K,j_K) \rbrace$, the net payoff to the attacker is: 
\begin{eqnarray}
J=\sum_{l=1}^{k-1} S_l^{i,j}(\beta_l)+I(i_K,j_K)S_K^{i_K,j_K}
\end{eqnarray} The second term $I(i_K,j_K)S_K^{i_K,j_K}$ indicates the cost at a SS. Furthermore, the proposed solution approach is dynamic in nature. The policy of the grid operator is a set of probability distributions $\mathcal{P}:=\lbrace p_1,\dots, p_K \rbrace\in\Delta_2^K$. Similarly,  for the attacker is a set of probability
distributions $\mathcal{Q}:=\lbrace q_1,\dots, q_K \rbrace\in\Delta_2^K$, where $\Delta_2$ is the  2-dimensional probability simplex. The expected pay-off of the game at any stage $s\in 1,2, \dots,K$ can be found using the forward recursive equation:
\begin{eqnarray}\label{rde}
J_s=\sum_{k=1}^s q_k^{\prime} \mathbf{S}_k p_k-\sum_{l=1}^{s-1}q_{k,1}p_{k,1}J_{s-l}.
\end{eqnarray} (\ref{rde}) is used in the next section to find the closed-form solution. A pair of policy $(\mathcal{P}^*,\mathcal{Q}^*)$ is the Nash equilibrium, if it ensures that none of the players can improve their respective payoffs by deviating from their policies, when the other players are employing the Nash equilibrium policy. This condition for Nash equilibrium policy is formally defined as:
$$J_K(\mathcal{P}^*,\mathcal{Q})\leq J_K(\mathcal{P}^*,\mathcal{Q}^*)\leq J_K(\mathcal{P},\mathcal{Q}^*),\forall \mathcal{P},\mathcal{Q}\in \Delta_2^K.$$ The outcome of the game, i.e., $J_K^*:=J_K(\mathcal{P}^*,\mathcal{Q}^*)$, is solved in the next section.

In the first level, the attacker starts the direct attack on a random combination of PMUs, denoted by $\lbrace \alpha_i\rbrace$. On the other hand, the grid operator starts with a initial PMU placement of $\lbrace y_i\rbrace$, and thereby reorganized as per the attack strategy. However, there can be a significant number of combination of actions for the attackers and grid operator in the first level game. In the second level game, the attacker selects the optimal PMU combination to attack directly in first level, whereas the grid operator chooses the optimal series of PMU placement strategy. The possible combinations of direct and indirect attack scenarios, denoted as $\lbrace A_i\rbrace$, consist of all possible paths from the root node to the leaf nodes in Fig. \ref{game_model}. All possible combinations of PMUs placement are denoted as $\lbrace y_i\rbrace$. Thus, the action sets for the attacker and the grid operator for the second level game are denoted as $\lbrace A_i\rbrace$ and $\lbrace y_i\rbrace$, respectively. The objective of the grid operator is to choose those actions of the first level with the minimum placement requirement of PMUs, while the attacker aims to select those actions that require more PMUs. 
This can be formally represented as:
\begin{eqnarray}\label{wsd}
\min_{\hat{p}\in \Delta_{|\lbrace y_i\rbrace|}} \max_{\hat{q}\in \Delta_{|\lbrace A_i\rbrace|}}\sum_{i\in \Delta_{|\lbrace y_i\rbrace|}}\sum_{j\in \Delta_{|\lbrace A_i\rbrace|}}\hat{q}_i\hat{p_j}(q_ip_j|\lbrace A_i\rbrace|)
\end{eqnarray} There always exist a solution of (\ref{wsd}) that satisfies the aforementioned Nash equilibrium conditions \cite{10.2307/2999512}.

\section{Solution Approach}
It can be observed from the first level of the game model in Fig. \ref{game_model}, that the game stops either in the
SS or at any of the other nodes in stage $K$. 
The value of the game, i.e., the expected payoff obtained by the attacker, at any stage can be represented in terms of the value at the next stage and payoff at the current stage. This is captured by the recursive equation: 
\begin{eqnarray}
V_{k-1}=\text{Val}\bigg(\begin{bmatrix}V_{k}^{11}&V_{k}^{12}\\V_{k}^{21}&V_{k}^{22}\end{bmatrix}+\begin{bmatrix}s^{11}_{k-1}&s^{12}_{k-1}\\s^{21}_{k-1}&s^{22}_{k-1}\end{bmatrix}\bigg), 
\end{eqnarray} where the stage $k\in\lbrace K,K-1,\dots,1\rbrace$. 
The expected value of the game at stage $k$ is represented by $V_{k}^{ij}$, where $i$ and $j$ denotes the action of the attacker and the grid operator. The exact payoff in stage $k-1$ is denoted as $S$, consisting of $s^{11}_{k-1}$, $s^{12}_{k-1}$, $s^{21}_{k-1}$, and $s^{22}_{k-1}$ as the payoffs of various actions of the players. $\text{Val}(.)$ is a mapping function which takes as inputs the expected value of the next stage $k$ and stage cost,  and returns the expected cost stage $k-1$.
\begin{IEEEeqnarray}{lCr}
V_{k-1}
=\max_{p_k\in \Delta_2} \min_{q_k\in \Delta_2}p_k^{\prime}\bigg(\begin{bmatrix}V_{k}^{11}&V_{k}^{12}\\V_{k}^{21}&V_{k}^{22}\end{bmatrix}+\begin{bmatrix}s^{11}_{k-1}&s^{12}_{k-1}\\s^{21}_{k-1}&s^{22}_{k-1}\end{bmatrix}\bigg)q_k\nonumber\\
\end{IEEEeqnarray}
The expected value of the game for a given stage at the Nash equilibrium strategy $\lbrace p_k^*, q_k^*\rbrace$ is given as:
\begin{eqnarray}
V_{k-1}=p_k^{*\prime}\bigg(\begin{bmatrix}V_{k}^{11}&V_{k}^{12}\\V_{k}^{21}&V_{k}^{22}\end{bmatrix}+\begin{bmatrix}s^{11}_{k-1}&s^{12}_{k-1}\\s^{21}_{k-1}&s^{22}_{k-1}\end{bmatrix}\bigg)q_k^*
\end{eqnarray}

\textbf{\textit{Policy of the Grid Operator:}} The expected value of the game at any stage $k$ from the perspective of the grid operator is given as:
\begin{IEEEeqnarray}{lCr}
V_k=\max_{ p\in\lbrace p_1, p_2\rbrace}\min_{q\in\lbrace q_1, q_2\rbrace}(p_1(V_{k}^{11}+s^{11}_{k-1})+p_2(V_{k}^{21}+s^{21}_{k-1}))q_1\nonumber\\+(p_1(V_{k}^{12}+s^{12}_{k-1})+p_2(V_{k}^{22}+s^{22}_{k-1}))q_2\nonumber\\
\end{IEEEeqnarray} Using the 2-dimensional probability simplex, we get $p_2=1-p_1$. This leads to the following: 
\begin{IEEEeqnarray}{lCr}\label{1}
V_k=\nonumber\\\max_{p_1\in[0,1]}\min \hspace{-0.2em}\begin{bmatrix}p_1(V_{k}^{11}+s^{11}_{k-1})\hspace{-0.2em}+\hspace{-0.2em}(1-p_1)(V_{k}^{21}+s^{21}_{k-1})\\p_1(V_{k}^{12}+s^{12}_{k-1})\hspace{-0.2em}+\hspace{-0.2em}(1-p_1)(V_{k}^{22}+s^{22}_{k-1})\end{bmatrix}
\end{IEEEeqnarray}

Given the policy of the attacker, the expected outcome over any action for the grid operator must be
the equal to each other. Applying this concept in (\ref{1}), the policy of grid operator is found to be:
\begin{eqnarray}
p_1(V_{k}^{11}+s^{11}_{k-1})+(V_{k}^{21}+s^{21}_{k-1})-p_1(V_{k}^{21}+s^{21}_{k-1})\nonumber\\=p_1(V_{k}^{12}+s^{12}_{k-1})+(V_{k}^{22}+s^{22}_{k-1})-p_1(V_{k}^{22}+s^{22}_{k-1})
\end{eqnarray}
\vspace{-0.1in}
\begin{IEEEeqnarray}{lCr}\label{optp}
\implies p_1^*=\nonumber\\\frac{(V_{k}^{22}+s^{22}_{k-1})-(V_{k}^{21}+s^{21}_{k-1})}{((V_{k}^{11}\hspace{-0.2em}+\hspace{-0.2em}s^{11}_{k-1})\hspace{-0.2em}-\hspace{-0.2em}(V_{k}^{21}\hspace{-0.2em}+\hspace{-0.2em}s^{21}_{k-1})\hspace{-0.2em}-\hspace{-0.2em}(V_{k}^{12}\hspace{-0.2em}+\hspace{-0.2em}s^{12}_{k-1})\hspace{-0.2em}+\hspace{-0.2em}(V_{k}^{22}\hspace{-0.2em}+\hspace{-0.2em}s^{22}_{k-1}))}\nonumber\\&
\end{IEEEeqnarray}

The probability of choosing the alternate action is given as: 
\begin{IEEEeqnarray}{lCr}
\implies p_2^*=\nonumber\\\frac{(V_{k}^{11}+s^{11}_{k-1})-(V_{k}^{12}+s^{12}_{k-1})}{((V_{k}^{11}\hspace{-0.2em}+\hspace{-0.2em}s^{11}_{k-1})\hspace{-0.2em}-\hspace{-0.2em}(V_{k}^{21}\hspace{-0.2em}+\hspace{-0.2em}s^{21}_{k-1})\hspace{-0.2em}-\hspace{-0.2em}(V_{k}^{12}\hspace{-0.2em}+\hspace{-0.2em}s^{12}_{k-1})\hspace{-0.2em}+\hspace{-0.2em}(V_{k}^{22}\hspace{-0.2em}+\hspace{-0.2em}s^{22}_{k-1}))}\nonumber\\
\end{IEEEeqnarray}

\textbf{\textit{Policy of the Attacker:}} The expected value of a game at any stage $k$ from the perspective of the attacker is given as:
\begin{IEEEeqnarray}{lCr}\label{2}
v_k=\nonumber\\\max\min_{q_1\in[0,1]}\hspace{-0.2em} \begin{bmatrix}(V_{k}^{11}+s^{11}_{k-1})q_1\hspace{-0.2em}+\hspace{-0.2em}(V_{k}^{12}+s^{12}_{k-1})(1-q_1)\\(V_{k}^{21}+s^{21}_{k-1})q_1\hspace{-0.2em}+\hspace{-0.2em}(V_{k}^{22}+s^{22}_{k-1})(1-q_1)\end{bmatrix}
\end{IEEEeqnarray}Given the policy of the grid operator, the expected outcome for the attacker over any action must be
the equal to each other. Applying this approach in (\ref{1}), the policy of the attacker is found to be:
\begin{eqnarray}
V_{k}^{11}+s^{11}_{k-1})q_1+(V_{k}^{12}+s^{12}_{k-1})-(V_{k}^{12}+s^{12}_{k-1})q_1\nonumber\\=(V_{k}^{21}+s^{21}_{k-1})q_1+(V_{k}^{22}+s^{22}_{k-1})-(V_{k}^{22}+s^{22}_{k-1})q_1
\end{eqnarray}
\begin{IEEEeqnarray}{lCr}\label{optq}
\implies q_1^*=\nonumber\\\frac{(V_{k}^{22}+s^{22}_{k-1})-(V_{k}^{12}+s^{12}_{k-1})}{((V_{k}^{11}\hspace{-0.2em}+\hspace{-0.2em}s^{11}_{k-1})\hspace{-0.2em}-\hspace{-0.2em}(V_{k}^{21}\hspace{-0.2em}+\hspace{-0.2em}s^{21}_{k-1})\hspace{-0.2em}-\hspace{-0.2em}(V_{k}^{12}\hspace{-0.2em}+\hspace{-0.2em}s^{12}_{k-1})\hspace{-0.2em}+\hspace{-0.2em}(V_{k}^{22}\hspace{-0.2em}+\hspace{-0.2em}s^{22}_{k-1}))}\nonumber\\
\end{IEEEeqnarray}On the other hand, the probability of selection the ``No Attack'' action is given as: 
\begin{IEEEeqnarray}{lCr}
\implies q_2^*=\nonumber\\\frac{(V_{k}^{11}+s^{11}_{k-1})-(V_{k}^{21}+s^{21}_{k-1})}{((V_{k}^{11}\hspace{-0.2em}+\hspace{-0.2em}s^{11}_{k-1})\hspace{-0.2em}-\hspace{-0.2em}(V_{k}^{21}\hspace{-0.2em}+\hspace{-0.2em}s^{21}_{k-1})\hspace{-0.2em}-\hspace{-0.2em}(V_{k}^{12}\hspace{-0.2em}+\hspace{-0.2em}s^{12}_{k-1})\hspace{-0.2em}+\hspace{-0.2em}(V_{k}^{22}\hspace{-0.2em}+\hspace{-0.2em}s^{22}_{k-1}))}\nonumber\\
\end{IEEEeqnarray}

\textbf{\textit{Value of the game:}} Given that the grid operator and the attacker employs the policies in (\ref{optp}) and (\ref{optq}), respectively, the value of the game at any stage $k$ is given by: 
\begin{IEEEeqnarray}{lCr}
V_{k-1}\nonumber\\=\frac{(V_{k}^{11}+s^{11}_{k-1})(V_{k}^{22}+s^{22}_{k-1})-(V_{k}^{12}+s^{12}_{k-1})(V_{k}^{21}+s^{21}_{k-1})}{s^{11}_{k-1}-s^{21}_{k-1}-s^{12}_{k-1}+s^{22}_{k-1}-V_k}\nonumber\\
\end{IEEEeqnarray} which can compactly represented as:
\begin{eqnarray}
V_{k-1}=V_k+\frac{s^{11}_{k-1}s^{22}_{k-1}-s^{12}_{k-1}s^{21}_{k-1}-s^{22}_{k-1}V_k}{s^{11}_{k-1}-s^{21}_{k-1}-s^{12}_{k-1}+s^{22}_{k-1}-V_k}
\end{eqnarray}
\vspace{-0.1in}
\section{Case Study}
     \begin{figure}[!t]
         \centering
         \vspace{-0.1in}
         \includegraphics[width=0.4\textwidth]{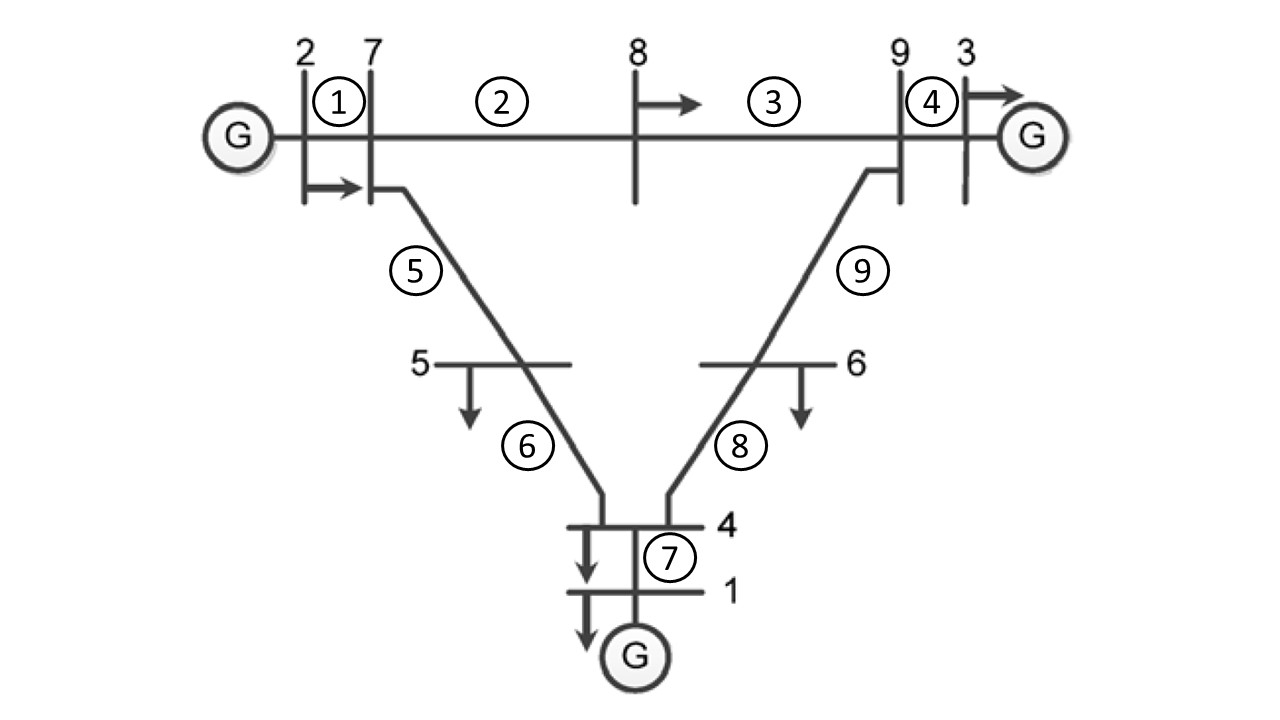}
         \vspace{-0.06in}
        \caption{IEEE 9-bus test system.}\label{IEEE9bus}
        \vspace{-0.3in}
     \end{figure}
\begin{table*}[!t]
\caption{Risk propagation probability matrix for 9-bus test system.} 
\vspace{-0.05in}
\centering 
\begin{tabular}{||c | c | c | c | c | c | c | c | c | c ||} 
\hline \hline
\#Bus/PMUs & $1$ & $2$ & $3$ & $4$ & $5$ & $6$ & $7$ & $8$ & $9$\\ 
\hline  \hline
$1$ & 1 & 0.002 & 0.001 & 0.001 & 0.002 & 0.001 & 0.002 & 0.001 & 0.001\\ \hline 
$2$ & 0.002 & 1 & 0.001 & 0.002 & 0.002 & 0.001 & 0.001 & 0.002 & 0.002\\ \hline 
$3$ & 0.001 & 0.001 & 1 & 0.002 & 0.002 & 0.001 & 0.002 & 0.001 & 0.001\\ \hline 
$4$ & 0.001 & 0.002 & 0.002 & 1 & 0.002 & 0.001 & 0.002 & 0.001 & 0.002\\ \hline 
$5$ & 0.002 & 0.002 & 0.002 & 0.002 & 1 & 0.002 & 0.002 & 0.001 & 0.001\\ \hline 
$6$ & 0.001 & 0.001 & 0.002 & 0.001 & 0.002 & 1 & 0.001 & 0.002 & 0.002\\ \hline 
$7$ & 0.002 & 0.001 & 0.002 & 0.002 & 0.002 & 0.001 & 1 & 0.002 & 0.001\\ \hline 
$8$ & 0.001 & 0.002 & 0.001 & 0.001 & 0.001 & 0.002 & 0.002 & 1 & 0.002\\ \hline 
$9$ & 0.001 & 0.002 & 0.001 & 0.002 & 0.001 & 0.002 & 0.001 & 0.002 & 1\\ 
\hline \hline
\end{tabular}
\vspace{-0.15in}
\label{wt}
\end{table*}

A modified IEEE 9-bus System (shown in Fig. \ref{IEEE9bus}) is used for this study. It consists of three generators connected to nine buses and seven buses are connected to loads. The current measurement incidence matrix for the  system is denoted as $A$. Rows correspond to PMU current measurements, and columns correspond to buses fitted with PMUs. The start and the end of the line being measured is denoted by 1 and -1, respectively.
     \vspace{-0.1in}

     {\small\begin{eqnarray}\label{ew}
     \small
     A=\begin{bmatrix} 
     0 & 1 & 0 & 0 & 0 & 0 & -1 & 0 & 0\\
     0 & -1 & 0 & 0 & 0 & 0 & 1 & 0 & 0\\
     0 & 0 & 0 & 0 & 0 & 0 & 1 & -1 & 0\\
     0 & 0 & 0 & 0 & 0 & 0 & -1 & 1 & 0\\
     0 & 0 & 0 & 0 & 0 & 0 & 0 & 1 & -1\\
     0 & 0 & 0 & 0 & 0 & 0 & 0 & -1 & 1\\
     0 & 0 & -1 & 0 & 0 & 0 & 0 & 0 & 1\\ 
     0 & 0 & 1 & 0 & 0 & 0 & 0 & 0 & -1\\
     0 & 0 & 0 & 0 & 0 & -1 & 0 & 0 & 1\\
     0 & 0 & 0 & 0 & 0 & 1 & 0 & 0 & -1\\
     0 & 0 & 0 & -1 & 0 & 1 & 0 & 0 & 0\\
     0 & 0 & 0 & 1 & 0 & -1 & 0 & 0 & 0\\
     -1 & 0 & 0 & 1 & 0 & 0 & 0 & 0 & 0\\
     1 & 0 & 0 & -1 & 0 & 0 & 0 & 0 & 0\\
     0 & 0 & 0 & 1 & -1 & 0 & 0 & 0 & 0\\
     0 & 0 & 0 & -1 & 1 & 0 & 0 & 0 & 0\\
     0 & 0 & 0 & 0 & 1 & 0 & -1 & 0 & 0\\
     0 & 0 & 0 & 0 & -1 & 0 & 1 & 0 & 0\\
     \end{bmatrix}
     \normalsize
     \end{eqnarray}}
     
     The diagonal matrix of series admittances of measured branches is denoted as:
     \vspace{-0.05in}
     \begin{eqnarray}\label{qy}
     Y=\text{diag}(y_1,\dots,y_9)
     \end{eqnarray}Utilizing (\ref{ew}) and (\ref{qy}), we obtain the $H_l$. Given that the attacker launches direct attack on a PMU installed on bus \#$i$, the probability of indirect attack $\beta_{ij}$ on the other PMUs is given Table \ref{wt}. Initially, the grid operator places the PMUs that ensure full system observability at the lowest cost, without considering the presence of PMUs. PMUs are placed at bus \#4 and \#7, i.e., 
     $$\lbrace y_i\rbrace=\lbrace 0,0,0,1,0,0,1,0,0\rbrace.$$ 
     
     Given the initial PMU placements, there can be four possible combinations of direct attacks launched, i.e.,
     $$\lbrace \alpha_i\rbrace=\lbrace\lbrace 0,0,0,1,0,0,0,0,0 \rbrace,\lbrace 0,0,0,0,0,0,1,0,0 \rbrace,$$
     $$\lbrace 0,0,0,1,0,0,1,0,0 \rbrace, \lbrace 0,0,0,0,0,0,0,0,0 \rbrace \rbrace$$
     Let us consider the attacker launches direct attack on PMU placed on bus \#4. Now using the row 4 of Table \ref{wt}, we can find the probability that the healthy PMU at bus \#7 can also get compromised, i.e., $\beta_{47}=0.002$. The attacker moves first by either launching false data injection attack or wait for the risk to propagate to other PMUs \cite{tian2022datadriven}. The goal of launching false data injection attacks is to maximize the state error covariance. This is achieved by solving (\ref{opt}). The optimized value of (\ref{me}) is: 
     $$\phi_D(\lbrace 0,0,0,1,0,0,1,0,0\rbrace)=30.2133.$$ In case the attacker chooses not to attack in the the first stage, as shown in Fig. \ref{game_model}, and launches attack in the second stage, then the optimized value of the error covariance metric is: $$\phi_D(\lbrace 0,0,0,1,0,0,1,0,0\rbrace)=32.43.$$
     \begin{figure}[!t]
         \centering
         \vspace{-0.1in}
         \includegraphics[width=0.35\textwidth]{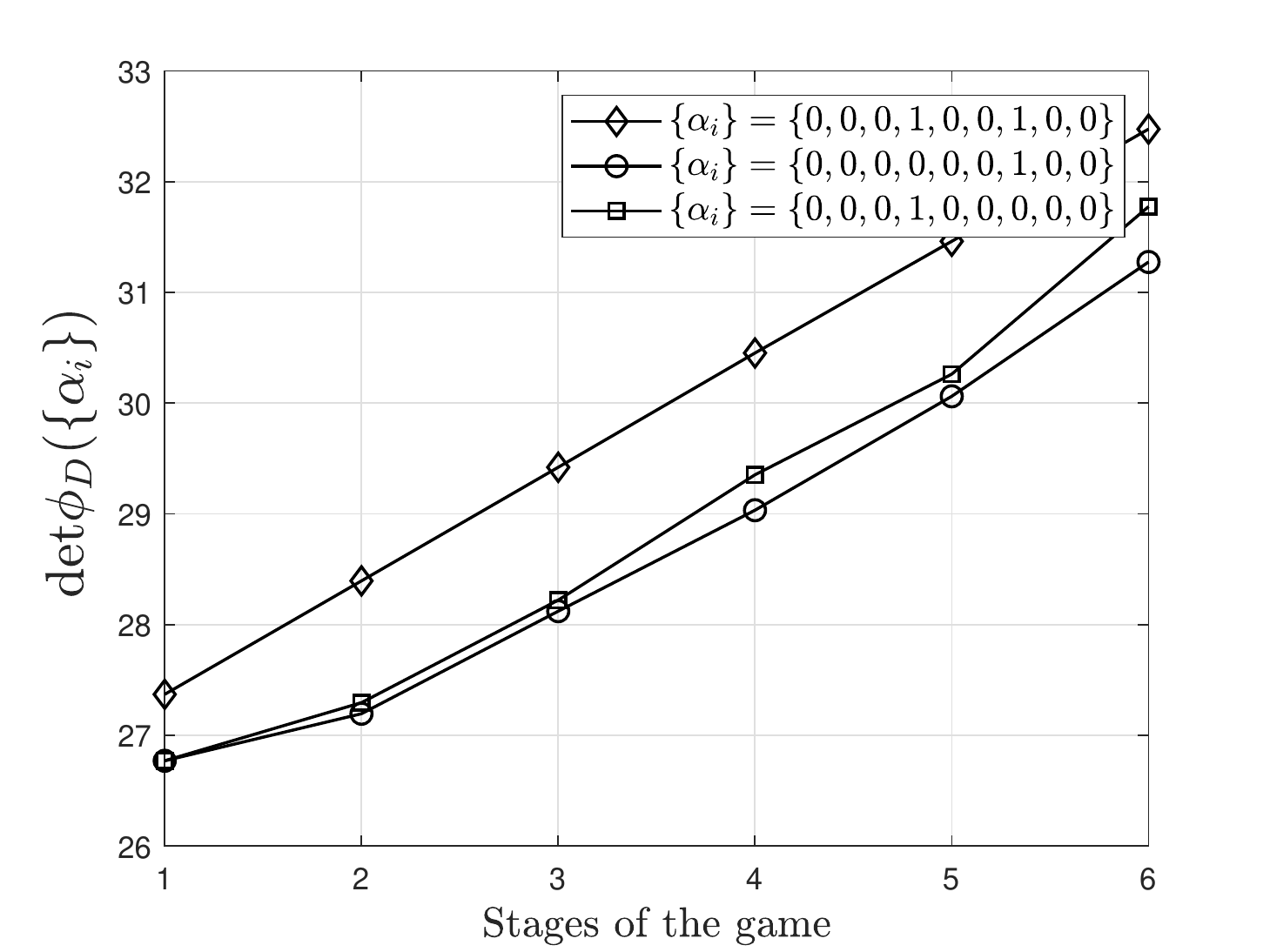}
         \vspace{-0.06in}
         \caption{Number of PMUs deployed in the IEEE 9-bus system.}\label{fig3}
         \vspace{-0.14in}
     \end{figure}
     \begin{figure}[!t]
         \centering
         \vspace{-0.1in}
         \includegraphics[width=0.4\textwidth]{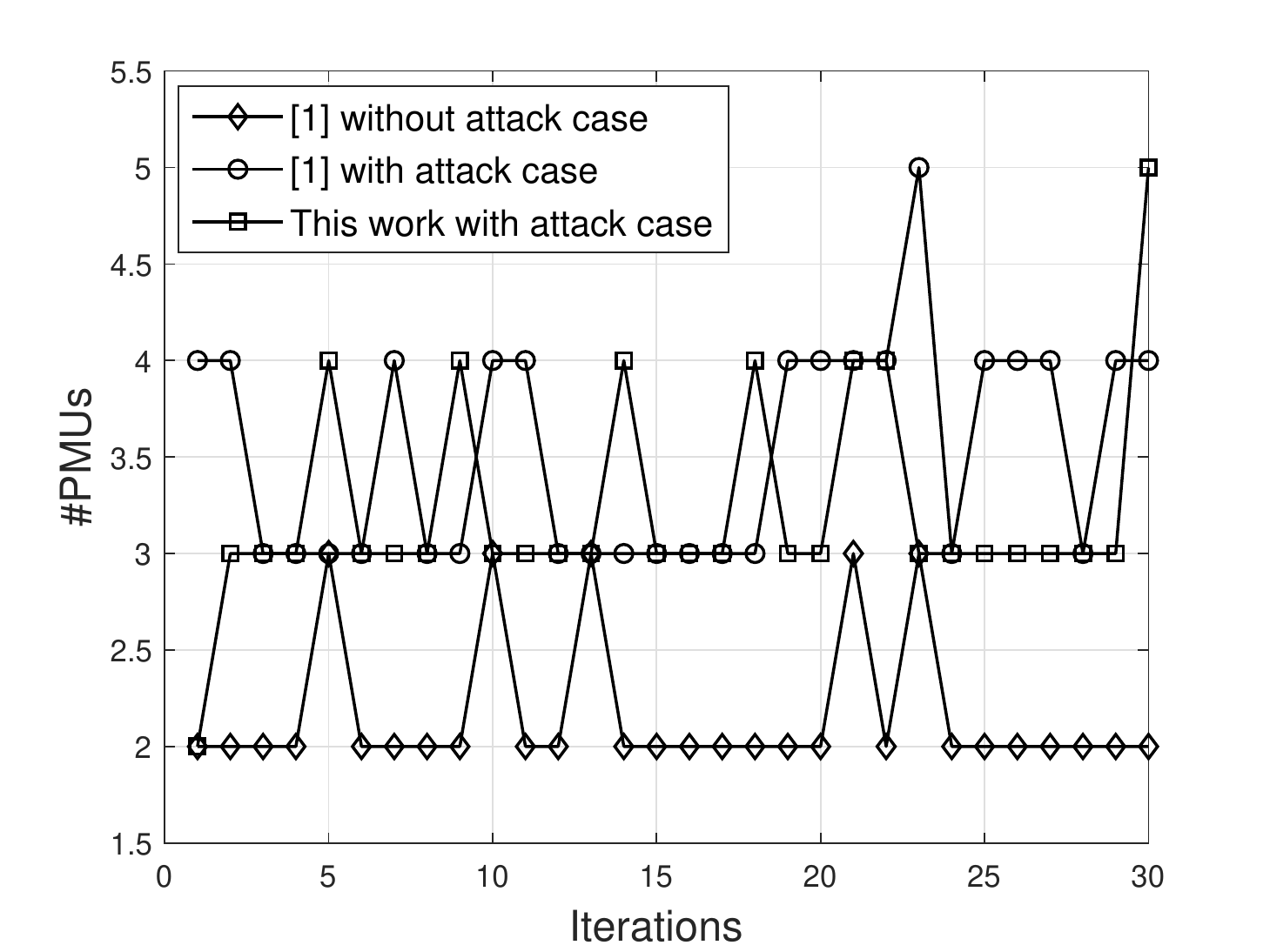}
         \vspace{-0.06in}
         \caption{Number of PMUs deployed in the IEEE 9-bus system.}\label{fig4}
         \vspace{-0.3in}
     \end{figure}
     \begin{table*}
     \caption{Average PMUs deployed for the compared scenarios.} 
     \vspace{-0.1in}
\centering
     \begin{tabular}{|| c | c ||} 
\hline \hline 
Scenarios& Average PMUs Deployed\\  
\hline \hline
\cite{8973553} without attack case & 2.33\\ \hline 
\cite{8973553} with attack case& 3.66 \\ \hline
This work with attack case & 3.33  \\  
\hline \hline
\end{tabular}
\vspace{-0.2in}
\label{wt1}
\end{table*}
     Fig. \ref{fig3} shows the plot of the error covariance metric with varying stages in which the false data injection is launched.  As the attacker delays the  data injection attacks, the risk propagation happens, i.e., a higher  number of healthy PMUs gets compromised from the direct attacks. Consequently, the attacker can launch the attacks in a larger pool of infected PMUs, increasing the state estimation error covariance metric. The direct attack strategy of $(\lbrace 0,0,0,1,0,0,1,0,0\rbrace)$ causes the most  damage in terms of error covariance metric than the other strategies. In the aforementioned attack, the attacker launches the direct attack on all the PMUs deployed by the grid operator, whereas in the other available strategies, the attacker directly attacks a subset of the deployed PMUs. Therefore, in the direct attack scheme, the attacker has access to more measurements to inject false data, and hence, the rate of risk propagation is also rapid.
     
     Based on the results of the first stage game, the second stage game is solved to choose the optimal strategy of the defender (resp. attacker) to minimize (maximize) the incurred cost for PMU deployment. The optimal strategies of the grid operator and the attacker obtained by solving the minmax problem in (\ref{opt}) is obtained as:
     \vspace{-0.05in}
     $$\lbrace y_i\rbrace^*=\lbrace 0.02,	0.0,	0.09,	0.23,	0.0,	0.15,	0.36,	0.0,	0.14\rbrace$$\vspace{-0.2in}
     $$\lbrace \alpha_i\rbrace^*=\lbrace 0.0,	0.013,	0.09,	0.25,	0.08,	0.08,	0.3,	0.0,	0.17\rbrace$$ 
     The exact cost incurred by the grid operator with the aforementioned strategies is plotted in Fig. \ref{fig4}. The average number of PMUs installed for the scenarios considered in Fig. \ref{fig4} is given in Table \ref{wt1}. Comparisons are made with the conventional PMU placement problem in \cite{8973553}, which achieves full observability of the buses using a minimum number of PMUs. On average, in \cite{8973553}, 2.33 PMUs were deployed in the network in an attack free scenario. However the required PMUs to maintain the observability constraints increased to 3.66 when the attack scenario was implemented. However, the proposed bi-level defense technique reduced the required number of PMUs to an average of 3.33, i.e., a performance improvement of 9\% in terms of deployment cost.

\vspace{-0.15in}

\section{Conclusions}
In this paper, the optimal PMU placement problem is investigated when a rational attacker is trying to compromise the deployed PMUs with the aim of injecting false data. This attack scenario affects  the cost of power system monitoring in terms of  observability  and error in the state estimation process. The proposed bi-level non-cooperative game model addresses the  state estimation error and the observability constraints in the first level, and then optimizes the cost in the second level. The technique is applied on the IEEE 9-bus system and reduces the PMU deployment cost when compared to traditional deployment algorithms.

\vspace{-0.15in}
\section*{Appendix -- Notation}\label{s:notation}
The following notions and conventions are employed throughout the paper:
$\mathbb{R},\mathbb{R}^m,\mathbb{R}^{m\times n}$  denote the space of real numbers, real vectors of length $m$ and real matrices of $m$ rows and $n$ columns respectively.
$X^\top$ denotes the transpose of the quantity $X$.
Normal-face lower-case letters ($x\in\mathbb{R}$) are used to represent real scalars, bold-face lower-case letter ($\mathbf{x}\in\mathbb{R}^m$) represents vectors, normal-face upper case ($X\in\mathbb{R}^{m\times n}$) represents matrices, while calligraphic upper case letters (e.g $\mathcal{T}$) represent sets. Let $\mathcal{T}\subseteq\{1,\hdots,m\}$ then, for a matrix $X\in\mathbb{R}^{m\times n}$, $X_\mathcal{T} \in\mathbb{R}^{\abs{\mathcal{T}}\times n}$ and $X^{\mathcal{T}} \in\mathbb{R}^{m\times \abs{\mathcal{T}}}$ are the sub-matrices obtained by extracting the rows, and columns respectively, of $X$ corresponding to the indices in $\mathcal{T}$. 
For a vector $\mathbf{x}$, $\mathbf{x}_i$ denotes its $i$th element. 
\vspace{-0.1in}
\section*{Acknowledgement}
This paper is an outcome of a larger 2-year project to develop intelligent solutions for grid technologies for ENOWA (NEOM) energy systems funded by ENOWA (NEOM) through a technical consultancy services agreement with KAUST.
\vspace{-0.1in}
\bibliographystyle{ieeetr}
\bibliography{ref1}

\end{document}